\documentstyle[12pt]{article}
\renewcommand{\a}{\alpha}

\newcommand{\g}{\gamma}           
\renewcommand{\d}{\delta}         \newcommand{\D}{\Delta}

\newcommand{\ld}{\lambda}        
         
\newcommand{\p}{\psi}              
\newcommand{\ro}{\rho}
\newcommand{\s}{\sigma}           
\newcommand{\th}{\theta}         
\newcommand{\f}{{\phi}}           
\newcommand{\vf}{{\varphi}}
       
\newcommand{\z}{\zeta} 
 
\newcommand{\cH}{{\cal H}}

\newcommand{\cL}{{\cal L}}
\newcommand{\cM}{{\cal M}}

\newcommand{\cO}{{\cal O}}

\newcommand{\cS}{{\cal S}}

\newcommand{\cU}{{\cal U}}

\newcommand{\cX}{{\cal X}}

\newcommand{\cZ}{{\cal Z}}

\newcommand{\hU}{{\widehat U}}

\newcommand{\Hx}{{\widehat x}}

\newcommand{\bz}{{\bar{z}}}
\newcommand{\bv}{{\bar{v}}}

\newcommand{\bt}{{\bar{t}}}
\newcommand{\bpa}{{\bar{\pa}}}
\newcommand{\bxi}{{\bar{\xi}}}

\newcommand{\bN}{\bar{N}}
\newcommand{\deff}{\,\stackrel{\rm def}{\equiv}\,}
\newcommand{\lra}{\longrightarrow}
\newcommand{\ra}{\,\rightarrow\,}
\def\limar#1#2{\,\raise0.3ex\hbox{$\longrightarrow$\kern-1.5em\raise-1.1ex
\hbox{$\scriptstyle{#1\rightarrow #2}$}}\,}
\def\limarr#1#2{\,\raise0.3ex\hbox{$\longrightarrow$\kern-1.5em\raise-1.3ex
\hbox{$\scriptstyle{#1\rightarrow #2}$}}\,}
\def\limlar#1#2{\ \raise0.3ex
\hbox{$-\hspace{-0.5em}-\hspace{-0.5em}-\hspace{-0.5em}
\longrightarrow$\kern-2.7em\raise-1.1ex
\hbox{$\scriptstyle{#1\rightarrow #2}$}}\ \ }

\newcommand{\wt}{\widetilde}
\newcommand{\os}{{\otimes}}
\newcommand{\da}{{\dagger}}
\newcommand{\stimes}{\times\hspace{-1.1 em}\supset}

\def\h{\hbar}

\newcommand{\exx}[1]{\exp\left\{ {#1}\right\}}

\newcommand{\one}{{\leavevmode{\rm 1\mkern -5.4mu I}}}

\newcommand{\Ibb}[1]{ {\rm I\ifmmode\mkern
            -3.6mu\else\kern -.2em\fi#1}}
\newcommand{\ibb}[1]{\leavevmode\hbox{\kern.3em\vrule
     height 1.2ex depth -.3ex width .2pt\kern-.3em\rm#1}}

\newcommand{\C}{{\ibb C}}
\newcommand{\R}{{\Ibb R}}

\newcommand{\rational}{{\kern .1em {\raise .47ex
\hbox{$\scripscriptstyle |$}}
    \kern -.35em {\rm Q}}}
\newcommand{\bm}[1]{\mbox{\boldmath${#1}$}}

\newcommand{\LL}{\cL^2(\R^2)}
\newcommand{\LLS}{\cL^2(S)}

\newcommand{\er}{{\rm{e}}}
\renewcommand{\i}{{\rm{i}}}

\newcommand{\id}{{\rm{id}\,}}

\newcommand{\const}{{\rm{\,const\,}}}

\newcommand{\diag}{{\rm{\,diag\,}}}

\newcommand{\PF}{{\bm{\cZ}}}
\newcommand{\pa}{\partial}
\newcommand{\pad}[2]{{\frac{\partial #1}{\partial #2}}}

\newcommand{\eg}{{\em e.g.,\ }}
\newcommand{\ie}{{{\em i.e.,}\ }}
\newcommand{\etc}{{\em etc.\ }}

\newcommand{\cf}{{\em cf.\ }}
\def\<{\langle}
\def\>{\rangle}
\def\lgl{\langle\langle}
\def\rgr{\rangle\rangle}
\newcommand{\bra}[1]{\left\langle {#1}\right|}
\newcommand{\ket}[1]{\left| {#1}\right\rangle}

\newcommand{\be}{\begin{equation}}
\newcommand{\ee}{\end{equation}}
\newcommand{\bn}{\begin{eqnarray}}
\newcommand{\en}{\end{eqnarray}}
\newcommand{\bnn}{\begin{eqnarray*}}
\newcommand{\enn}{\end{eqnarray*}}
\newcommand{\ba}{\begin{array}}
\newcommand{\ea}{\end{array}}
\newcommand{\e}{\label}
\newcommand{\nbr}{\nonumber\\[3mm]}
\newcommand{\r}[1]{(\ref{#1})}

\newcommand{\qq}{\qquad}

\newcommand{\biz}{\begin{itemize}} 
\newcommand{\eiz}{\end{itemize}}
\newcommand{\ben}{\begin{enumerate}} 
\newcommand{\een}{\end{enumerate}}

\newcommand{\fr}{Fourier\ }

\def\raczka{a\kern-0.35em\raise-0.4ex\hbox{$\scriptscriptstyle c$}}

\newcommand{\se}{\sqrt{\eta}}
\newcommand{\paq}{\partial_q}
\newcommand{\bpaq}{\bar{\partial}_q}
\begin{document}
\begin{titlepage}
\
\vspace{0.5cm}
\begin{flushright}
{\bf HIP-1999-19/TH}\\
April 20, 1999
\end{flushright}

\begin{center}
\vspace*{1.0cm}

{\Large \bf Quantum Field Theory\\ 
\vspace{5mm}
on the Noncommutative Plane with $E_q(2)$ Symmetry}

\vskip 1cm

{\large {\bf M. Chaichian}}, 
\ \ \ {\large{\bf 
A. Demichev}}\renewcommand{\thefootnote}{a}
\footnote{Permanent address: 
Nuclear Physics Institute, Moscow State University,
119899, Moscow, Russia}\ \ and \ \ 
{\large{\bf 
P. Pre\v{s}najder}}\renewcommand{\thefootnote}{b}
\footnote{Permanent address:
Department of Theoretical Physics, Comenius University,
Mlynsk\'{a} dolina, SK-84215 Bratislava, Slovakia}

\vskip 0.2cm

High Energy Physics Division, Department of Physics,\\
University of Helsinki\\
\ \ {\small\it and}\\
\ \ Helsinki Institute of Physics,\\
P.O. Box 9, FIN-00014 Helsinki, Finland

\end{center}

\vspace{2 cm}

\begin{abstract}
\normalsize

We study properties of a scalar quantum field theory on the two-dimensional
noncommutative plane with $E_q(2)$ quantum symmetry. We start from the
consideration of a firstly quantized quantum particle on the noncommutative
plane. Then we define quantum fields depending on noncommutative coordinates
and construct a field theoretical action using the $E_q(2)$-invariant measure
on the noncommutative plane. With the help of the partial wave decomposition
we show that this quantum field theory can be considered as a second
quantization of the particle theory on the noncommutative plane and that this
field theory has (contrary to the common belief) even more severe ultraviolet
divergences than its counterpart on the usual commutative plane. Finally, we
introduce the symmetry transformations of physical states on noncommutative
spaces and discuss them in detail for the case of the $E_q(2)$ quantum group.

\end{abstract}

\vspace{0.5cm}


\end{titlepage}

\section{Introduction}

It is generally believed that the picture of space-time as a manifold $\cM$
should break down at very short distances of the order of the Planck length.
One possible approach to the description of physical phenomena at small
distances is based on noncommutative geometry of space-time. There have
been investigations in the context of Connes' approach \cite{Connes} to
gravity and the Standard Model of electroweak and strong interactions
\cite{ConnesL,Chams} and in the framework of the string theory 
\cite{strings-nc}. Another approach starting from study of a relation 
between measurements at very small distances and black hole formations 
has been developed in the pioneering works \cite{DoplicherFR}. 
One more possibility is based on Quantum Group theory 
(see, \eg \cite{ChaichianDbk}).

The essence of the noncommutative geometry consists in reformulating first
the geometry in terms of commutative algebras and modules of smooth
functions, and then generalizing them to their noncommutative analogs.  If
the notions of the noncommutative geometry are used directly for the
description of the space-time, the notion of points as elementary geometrical
entity is lost and one may expect that an ultraviolet cutoff appears.

As is well known from the standard quantum mechanics, a quantization of any
compact space, in particular a sphere, leads to finite-dimensional
representations of the corresponding operators, so that in this case any
calculation is reduced to manipulations with finite-dimensional matrices and
thus there is simply no place for UV-divergences (see \cite{Ber,Hoppe,GKP3}
and refs. therein).  Things are not so easy in the case of noncompact
manifolds. The quantization leads to infinite-dimensional representations and
we have no guarantee that noncommutativity of the space-time coordinates
removes UV-divergences.  In our preceding paper \cite{CDP98} we have shown
that ultraviolet behaviour of a field theory on a noncommutative space-time is
sensitive to the topology of the space-time, namely to its compactness. We
considered theories on a two-dimensional plane with Heisenberg-like
commutation relations among coordinates (see also \cite{DoplicherFR,Filk})
and on a noncommutative cylinder. While the former retains the divergent
tadpoles (as an ordinary QFT), the latter proves to be UV-finite.  We argued
that the underlying reason for such a UV-behaviour of the models is related
to the properties of the complete coordinate-momentum quantum mechanical
algebra and to the fact that the momenta degrees of freedom are associated to
the fully {\it noncompact} Heisenberg-Weyl group manifold in the first case
and to the cylinder in the second case (the cylinder has one {\it compact}
dimension).

Using these qualitative arguments, we supposed that the quantum field theory
constructed on the $q$-deformed plane \cite{VaksmanK,SchuppWZ,BonecciCGST}
with $E_q(2)$-symmetry also has
UV-divergences. We have proved that indeed there are no kinematical reasons
for this model to be UV-finite: the Green function of the free theory on the 
$q$-plane is singular. Moreover, we have shown that the interaction with an
external field does produce divergent tadpole. However, in the paper
\cite{CDP98} we used decomposition of the fields on the $q$-plane in the
so-called distorted plane waves ($q$-deformed exponential functions). This
makes difficult matching the $q$-deformed field theory with the corresponding
firstly quantized quantum mechanics of particles on the q-deformed plane and
due to the absence of the additivity property for the $q$-exponentials,
makes an explicit calculation of nontrivial (\eg $\vf^4$-)
vertices impossible. 
Thus the results of \cite{CDP98} have left open the possibility
that the complete interacting theory on the $q$-plane is UV-finite because of
(dynamical) properties of the corresponding $\vf^4$-vertices.

In this paper we use another decomposition of the fields, namely, the
decomposition in partial waves, similar to the recently proposed ``spherical
field theory'' \cite{Lee} on commutative spaces. This decomposition together
with the Haar ($E_q(2)$-invariant) measure and $q$-deformed integral used for
the definition of the field theoretical action, allows to present the field
theory on the $q$-deformed plane as a lattice theory of infinite number of
interacting one-dimensional fields (partial waves). The resulting field
theoretical degrees of freedom are in transparent correspondence with the
spectrum of operators in the firstly quantized version of the model. The
calculation of the tadpole with the account of the $\vf^4$-vertex shows that
UV-properties of the theory on the $q$-deformed plane are even worse than those
on the ordinary commutative plane. This fact confirms the conclusion of the
paper \cite{CDP98} that the very transition to the noncommutative space-times
does not guarantee UV-finiteness.

The example of the plane with the most simple and natural Heisenberg-like
commutation relations among coordinates was used in \cite{CDP98} also for
study of symmetry transformations of noncommutative space-times with Lie
algebra commutation relations for coordinates. The noncommutative coordinates
prove to be tensor operators, and we considered concrete examples of the
corresponding transformations of localized states (analog of space-time point
transformations). In this paper, we extend this consideration to the much
more involved case of quantum group coaction on noncommutative space-times.
More precisely, we derive the rules of transformations of particle states
induced by the coaction of a quantum group.

The paper is organized as follows. In section 2 we consider firstly quantized
theory of particles on the $q$-deformed plane $P^{(2)}_q$ with
$E_q(2)$-symmetry. We derive representations of the algebras of coordinates
and momenta on the $q$-plane and find spectra of the relevant operators.  In
section 3 the field theory (second quantization) on $P^{(2)}_q$ is introduced
and presented in the form of infinite number of interacting partial waves
defined on a one-dimensional lattice, the partial wave at the sites of the
lattice (interpreted as creation and annihilation operators) being in
one-to-one correspondence with spectra of the quantum mechanical operators
found in section 2. Calculation of a tadpole diagram shows that
the model has even more severe UV-divergences than the standard
two-dimensional scalar $\vf^4$-theory. In section 4 we are interested in
transformation properties of a system on $P^{(2)}_q$ under the coaction of
the quantum group $E_q(2)$. The point is that now the coordinates $\bz, z$
are noncommuting operators and $E_q(2)$ provides only existence of coaction,
\ie homomorphism of the algebra of functions on $P^{(2)}_q$ into the direct
product $E_q(2)\otimes P^{(2)}_q$ of algebras of functions on the quantum
group and plane. Then the question is: how does this coaction influence
states of a quantum system on $P^{(2)}_q$? In other words, if a system is in
some state $\p$ (say, with a definite value of one of the coordinate
operators, $z$ or $\bz$ or some their combination) we are interested in
determination of the state after the $E_q(2)$-group coaction. In subsection
4.1 we clarify a general formulation of this problem and then (in subsection
4.2) give an explicit answer for the $E_q(2)$ group. Section 5 is devoted to
the summary of the results.  

\section{Quantum mechanics on the noncommutative plane with quantum $E_q(2)$
group symmetry} 

In this paper we consider Quantum Mechanics induced by a quantum group 
structure.
Recall that in the case of ordinary Lie group $G$, the group structure defines
a unique symplectic structure on the cotangent bundle $T^*_G$ to the group
manifold $G$ (see, \eg \cite{LibermannM}) and, hence, the corresponding
canonical quantization (via substitution of Poisson brackets by the
corresponding commutators). A similar construction with necessary
generalizations, can be carried out for Lie-Poisson groups,
which after the quantization procedure become quantum groups (see, \eg review
in \cite{ChaichianDbk} and refs. therein). 

In fact, the quantization of a system on a Lie group cotangent bundle $T^*_G$
corresponds to choice of the group manifold as a configuration space (\ie
group parameters as space coordinates) and left- (or right-)invariant vector
fields on $G$ (elements of the corresponding Lie algebra) as quantum
mechanical momenta. Instead of using a whole group $G$, one can start from
some of its coset (homogeneous) space $G/H$, where $H$ is a subgroup
$H\subset G$. In this approach the basic problem of Quantum Mechanics, \ie
determination of possible representations of canonical operators, is reduced
to mathematically well-developed problem of construction of regular
representation (or quasi-regular, if one deals with a homogeneous space) and
its decomposition into irreducible parts (see, \eg \cite{BarutR}). In some
particular cases this general construction becomes rather simple and quite
familiar from elementary course on Quantum Mechanics. For example, let us
consider a two-dimensional Euclidean group $E(2)=U(1)\stimes T_2$ containing
rotations and translations of a two-dimensional plane. Its homogeneous space
$P^{(2)}=E(2)/U(1)$ is the Euclidean plane with the metric
$\eta_{ij}=\diag\{+1,+1\}$ which is invariant with respect to
$E(2)$-transformations. This configuration space is parameterized by two
coordinates $x_1, x_2$, while left-invariant fields tangent to this
homogeneous space are nothing but usual derivatives which up to the factor
$-\i\h$ correspond to the standard momentum operators. As is well-known, any
representation of the algebra of coordinates and left-invariant vector fields
on $P^{(2)}$ is unitary equivalent to this representation by the coordinate
functions and derivatives in the Hilbert space $\cH=\LL$ of square integrable
functions.  States $\p(x)\in\cH$ are transformed according to representations
of $E(2)$ in the Hilbert space $\cH$. Since the coordinate operators are
commuting, their eigenvalues are transformed under an action of $E(2)$-group
as their classical counterparts: in the convenient complex notation $$
z=x_1+\i x_2\ ,\qq \bar z=x_1-\i x_2\ , $$ the $E(2)$ transformations read
as
\bn
z\ra z'&=&vz+t\,\nonumber\\[-1mm]
&&\e{1.3}\\[-1mm]
\bar z\ra \bar z'&=&\bar v\bar z+\bar t\,\nonumber
\en
where $v,\ \bar v$ subjected to the constraint $\bar vv=1$, 
define two-dimensional rotation group $U(1)$ and $t=t_1+\i t_2\ ,
\bar t=t_1-\i t_2$ parameterize translations.

For the simple case of the quantum mechanical systems on the Euclidean plane
the underlying mathematics related to cotangent bundle structures, regular
representations \etc seems to be redundant. But for generalizations to more
complicated homogeneous spaces, in particular, with non-zero curvature and
nontrivial topology, the group theoretical methods become quite actual and
powerful.

In this work we are going to study another generalization: instead of
starting from ordinary $E(2)$, we shall use its quantum version $E_q(2)$ 
\cite{VaksmanK,SchuppWZ,BonecciCGST}. Though in the case of quantum groups
and corresponding quantum homogeneous spaces (definition of the latter see,
\eg in \cite{BonecciCGST}) group parameters (coordinates) become
noncommutative, the general scheme of quantization still can be applied. The
role of momentum operators is now attributed the $q$-deformed left- (or
right-) invariant 
generalizations of vector fields (see, \eg \cite{Zumino}). Thus the Planck
constant $\h$ enters, as usual, the commutation relations (CR) for momenta and
coordinates, while the group deformation parameter $q$ governs nontrivial
coordinate-coordinate and momentum-momentum CR. Therefore, first of all we
have to construct possible representations of this {\it combined} 
$q$-deformed algebra of noncommuting coordinates and momenta. 
For the particular case which we
consider in this paper ($q$-deformed quantum Euclidean plane $P^{(2)}_q$) this
is not a very complicated problem and we shall consider it in this section.

We start from the quantum group $E_q(2)$ generated by elements $\bv,\ v,\
\bt,\ t$ with the defining relations \cite{VaksmanK} 
\be\ba{lllll}
\bv v = v\bv = 1\ ,& \qq &t\bt = q^2 \bt t\ ,& \\[3mm]
vt= q^2tv\ ,& \qq &\bv t = q^{-2}t\bv &\qq& q\in\R\ .\ea       \e{2.1}
\ee
Other commutation relations follow from the involution: 
$v^\da =\bv,\ t^\da=\bt$. 
The comultiplication has the form
\begin{equation}
\begin{array}{cc}
\D v = v \otimes v\ , & \D\bv = 
\bv \otimes \bv\ , \\[2mm]
\D t = v \otimes t + t \otimes \one\ , &
\D\bt = \bv \otimes \bt + \bt \otimes {\one}\ .
\end{array}                                            \label{2.2}
\end{equation}
The explicit form of other basic maps for $E_q(2)$ (antipode, counity) will
not be used in what follows. 

The unitary element $v$ can be parameterized with the help of the symmetric
element $\th$:
\be\ba{lcl}
v=\er^{\i\th}\ ,&\qq&\th^\da=\th\ ,\\[2mm]
\ &\ &\D\th=\th\otimes\one+\one \otimes \th\ .\ea      \e{2.4}
\ee
The corresponding quantum universal enveloping algebra (QUEA) $\cU_qe(2)$ is
generated by the elements $J,\ \bar T,\ T$ which are dual to the generators
$\th,\ \bt,\ t$ of the algebra $E_q(2)$ and, as a result of the duality, 
satisfy the following commutation relations 
\be\ba{lll}
[J,T]=\i T\ ,&\qq& [J,\bar T]=-\i \bar T\ ,\\[2mm]
T\bar T=q^2\bar T T&&\ea                                 \e{2.6}
\ee
(comultiplication and the other basic maps are also defined by the duality).

The left action of elements from QUEA $\cU_q L$ of an arbitrary 
Lie algebra $L$ on
elements of the corresponding quantum group $G_q$ is defined by the
expressions
\be
\ell (X)f=(\id\otimes X)\cdot\D f\equiv\sum_if^i_{(1)}
\lgl X,f^i_{(2)}\rgr\ ,                                  \e{2.8}
\ee
or
\be
\ld (X)f=(S(X)\otimes \id)\cdot\D f\equiv\sum_i\lgl S(X),f^i_{(1)}\rgr
f^i_{(2)}\ ,                                  \e{2.8a}
\ee
where $X\in\cU_q L,\ f,f^i_{(1,2)}\in G_q$, $\lgl\cdot,\cdot\rgr$ denotes the
duality contraction, $S(X)$ is antipode and where the comultiplication 
in $G_q$ is presented in the
form $\D f= \sum_if^i_{(1)}\otimes f^i_{(2)}$. An explicit calculation of this
left action in the case of $E_q(2)$ shows that the operators
$\bar T,T\in\cU_qe(2)$ act on elements of $E_q(2)$ generated by $\bt,\ t$ 
exactly in the same way as the $q$-deformed derivatives $\bpaq,\ \paq$. 
In fact, the elements $\bt,\ t$ generate the $q$-deformed analog $P^{(2)}_q=
E_q(2)/U_q(1)$ of the homogeneous space $P^{(2)}$, \ie generate the algebra
of functions on quantum Euclidean plane \cite{BonecciCGST}. We shall denote
elements of the algebra $P^{(2)}_q$ by $\bz,\ z$ to distinguish them from
elements $\bt, \ t$ of the algebra  $E_q(2)$. 

The elements $\bz,\ z$ and $\bpaq,\ \paq$ defines the $q$-deformed algebra of
functions on $P^{(2)}_q$ together with the $q$-deformed left-invariant vector 
fields (derivatives). Its defining relations read as
\bn
&&z\bz=q^2\bz z\ ,\qq \pa_q\bpa_q=q^2\bpa_q\pa_q\nbr
&&\pa_q z=1+q^{-2}z\pa_q\ ,\qq\bpa_q \bz=1+q^{2}\bz\bpa_q\ ,\e{2.9}\\[3mm]
&&\bpa_q z=q^{2}z\bpa_q\ ,\qq\pa_q \bz=q^{-2}\bz\pa_q\ ,\nonumber
\en
(the commutation relation for the $q$-derivatives is just the rewritten
commutation relation for $\bar T,\ T$ \r{2.6} and
those for the $q$-derivatives and coordinates are derived from \r{2.8}). If we
put $q=1$ and define $p=-\i\h\pa,\ \bar p=-\i\h\bpa$, the relations \r{2.9}
become the usual canonical commutation relations for a particle in 
two-dimensional space.
The requirement of consistency with antipode dictates the following
conjugation rule for the $q$-derivatives \cite{SchuppWZ}
\be
\pa_q^\da=-q^2\bpa_q\ ,\qq\bpa_q^\da=-q^{-2}\pa_q\ .        \e{2.10}
\ee

We consider the relations \r{2.9} as a $q$-deformation of the 
canonical commutation relations and
is going to construct their representation in a Hilbert space.

To this aim let us introduce the operators $N$ and $\bN$ defined by the
relations
\be
[N;q^{-2}]=z\pa_q\ ,\qq [\bN;q^{2}]=\bz\bpa_q\ ,          \e{2.11}
\ee
\be
[X;q^{\a}]\equiv\frac{q^{\a X}-1}{q^{\a}-1}\ .  \e{q-number}
\ee
These operators have simple commutation relations 
\bn
&&q^{\a N}z=q^\a zq^{\a N}\ ,\qq q^{\a \bN}\bz=q^\a \bz q^{\a \bN}
\ ,\nonumber\\[-1mm]
&&          \e{2.15} \\[-1mm]
&&q^{\a N}\pa_q=q^{-\a}\pa_q q^{\a N}\ ,\qq
 q^{\a\bN}\bpa_q=q^{-\a}\bpa_q q^{\a\bN}\ .\nonumber
\en  
Using \r{2.10} and $z^\da=\bz$, we find
\be
N^\da=-\bN-1\ ,\qq \bN^\da=-N-1\ .                \e{2.16}
\ee 
The operators $q^{2\bN},\ q^{2N}$ allow to construct commuting pairs of
conjugate operators:
\bn
&&\bar Z= q^{N-\bN}\bz\ ,\qq Z= zq^{N-\bN}\ ,\nonumber\\[-1mm]
&&\e{2.20}\\[-1mm]
&&\bar P= qq^{-(N-\bN)}\bpa_q\ ,\qq P= -q^{-1}\pa_qq^{-(N-\bN)}\ ,\nonumber
\en
with the commutation relations
\bn
&&P\bar Z=\bar Z P\ ,\qq \bar Z Z=Z\bar Z\ ,\qq
 ZP=1+q^{2}PZ \ ,\nonumber\\[-1mm]
&&\e{2.22}\\[-1mm]
&&\bar P Z=Z\bar P\ ,\qq \bar P P=P\bar P\ ,\qq
\bar P\bar Z=1+q^{2}\bar Z\bar P\ .\nonumber
\en

If we were given only the algebra of the operators $\bz,\ z,\ \bpa_q,\ \pa_q$,
we would reasonably name the commuting operators $\bar Z,\ Z$ by coordinates
and $\bar P,\ P$ by the corresponding lattice momenta and then deal with two
independent (commuting with each other) one-dimensional algebras on the 
$q$-lattice. 
However, fields in NC-QFT depend on noncommutative ($q$-commuting)
coordinates $\bz,\ z$ which are more suitable to trace a
result of coaction by $E_q(2)$. We have found convenient to use the hermitian
and unitary combination of the coordinate operators:
\be
r^2\equiv z\bz\ \ \ \mbox{(hermitian)},\qq u\equiv\sqrt{\bz z^{-1}}
\ \ \ \mbox{(unitary)},                                 \e{2.37a}
\ee
together with $q^{(\bN-N)}$ (hermitian operator) and $q^{2(\bN+N+1)}$
(unitary operator) as a basic set of the phase space operators. The
commutation relations for this set of operators read as
\be\ba{lll}
[q^{2(\bN-N)},r^2]=0\ ,&\qq &[q^{(\bN+N+1)},u]=0\ ,\\[3mm]
r^2u=q^2u r^2\ , & \qq & [q^{2(\bN-N)},q^{(\bN+N+1)}]=0\ ,\\[3mm]
q^{2(\bN-N)}u=q^2u q^{2(\bN-N)}\ , & \qq &
q^{(\bN+N+1)}r^2=q^2r^2q^{(\bN+N+1)}\ ,\ea   \e{2.38}
\ee
Now we are ready to construct a representation of this algebra in the space
$\ell^2$ (\ie infinite dimensional matrix representation):
\bn
r^2\mid n,m\rangle_{r_0,l_0}  & = &  
r^2_0q^{2n}\mid n,m\rangle_{r_0,l_0} \nbr
q^{2(\bN-N)}\mid n,m\rangle_{r_0,l_0} & = 
& l_0q^{2m}\mid n,m\rangle_{r_0,l_0}\ .       \label{2.39}
\en
\bn
u\mid n,m\rangle_{r_0,l_0} & = & \mid n+1,m+1\rangle_{r_0,l_0}\ , \nbr
q^{(\bN+N+1)}\mid n,m\rangle_{r_0,l_0}& = &\mid n+1,m\rangle_{r_0,l_0}\ .
                                          \label{2.40}
\en
The constants $r_0$ and $l_0$ mark different representations and from
the eigenvalues of $r^2$ and $q^{2(\bN-N)}$ it follows that in the ranges
$[r_0,q^4 r_0)$ and $[l_0,q^4l_0)$ the representations are nonequivalent.
The matrices $r^2,q^{2(\bN-N)}$ are hermitian and $u,\ q^{(\bN+N+1)}$ 
are unitary with respect to the scalar product defined by
$$
{}_{r_0,l_0}\langle
n,m\mid n',m'\rangle_{r_0,l_0}=\delta_{nn'}\delta_{mm'}\ .      $$

Thus we have obtained that states of a particle on the quantum plane are
characterized by discrete values of its radius-vector and discrete values of
the operator $q^{2(\bN-N)}$ which is obviously related to deformation of the
angular momentum operator. Indeed, from \r{2.11} we conclude that the
operator 
\be
J_q\equiv[\bN-N;q^2]=\frac{q^{2(\bN-N)}-1}{q^{2}-1}\ ,   \e{2.41}
\ee
(which differs from $q^{2(\bN-N)}$ by multiplication and shifting by the
constants) in the continuum limit $q\ra1$ becomes the ordinary angular 
momentum operator. Therefore it is natural to consider $J_q$ as an appropriate
deformation of the latter. Of course, discreteness of values of an angular
momentum operator is not peculiar feature of $q$-deformed systems but general
property of all quantum systems.
Analogously, the natural $q$-deformation of the dilatation operator reads as
\be
D_q\equiv[(\bN+N+1);q^2]=\frac{q^{2(\bN+N+1)}-1}{q^{2}-1}\ ,   \e{2.42}
\ee

Another possibility for the construction of representations of the algebra
\r{2.38} which will be convenient for us in the next section is to
construct the representation in the basis of the unitary operators
\be
u\equiv\bz z^{-1}\ ,\qq\mbox{and}\qq q^{2(\bN+N+1)}\ , \e{2.43}
\ee
which commute with each other and, hence, have common eigenvalues. This basis,
certainly, less suitable for construction of a matrix representation of the kind
presented above since the two other (hermitian) operators do not shift an
eigenvector of the operators \r{2.43} exactly into another eigenvector. 
However, this basis proves to be more suitable for the study of
transformations of the states under the coaction of the quantum group $E_q(2)$
which we shall carry out in the section 4. 

\section{Quantum field theory on $P_q^{(2)}$ as a one-dimensional lattice
theory for an infinite set of interacting fields}

In this section we shall introduce the scalar $\vf^4$-field theory on the
noncommutative plane $P_q^{(2)}$ and present it in the form of infinite
number of interacting partial waves defined on a one-dimensional lattice.
 
\subsection{Preliminaries on ``the spherical field theory''}
The starting idea of the spherical field theory \cite{Lee} in a usual
commutative space-time is the
representation of a $d$-dimensional Euclidean field theory as a theory for an
infinite set of one-dimensional interacting fields. In what follows we shall
confine ourselves with the simplest case of two-dimensional scalar theory. The
initial action is quite standard:
\be
S=\int\,d^2x\,\left[\big(\pa_i\bar\vf\big)\big(\pa_i\vf\big)
+\mu^2\bar\vf\vf+\frac{\ld}{2}\big(\bar\vf\vf\big)^2
-j\bar\vf-\bar j\vf\right]\ .                                   \e{pw1}
\ee
Decomposing $\vf(x)$ and $j(x)$ into partial waves
\bn
\vf(x)&=&\vf(r,\a)=\frac{1}{\sqrt{2\pi}}\sum_{N=-\infty}^\infty\vf_N(r)
\er^{\i N\a}\ ,                                             \e{pw2}\\[3mm]
j(x)&=&j(r,\a)=\frac{1}{\sqrt{2\pi}}\sum_{N=-\infty}^\infty j_N(r)
\er^{-\i N\a}\ ,                                             \e{pw2a}
\en
one can rewrite \r{pw1} as
\bn
S&=&\sum_{N=-\infty}^\infty \int_0^\infty\,dr\left[r
\pad{\bar\vf_N}{r}\pad{\vf_N}{r}+\frac{\mu^2 r^2+N^2}{r}\bar\vf_N\vf_N
-rj_N\bar\vf_N-r\bar j_N\vf_N\right]\nbr
&+&\frac{\ld}{2}\sum_{N,M,K,L=-\infty}^\infty\int_0^\infty\,dr\,r\,
\Big(\bar\vf_N\vf_M\bar\vf_K\vf_L\,\d_{N-M+K-L,0}\Big)\ .     \e{pw3}
\en
Let $\wt{G}(k)$ denote the usual Green function in the momentum representation
\be
\wt{G}(k)=\int\,d^2x\,\er^{\i \vec{k}\vec{x}}\<0|\bar\vf(x)\vf(0)|0\>\ .\e{pw4}
\ee
Then the propagator for the $N$-th partial wave proves to be
\be
\<0|\bar\vf_N(r_1)\vf_N(r_2)|0\>=\int\,dk\,k\,J_{|N|}(kr_1)
J_{|N|}(kr_2)\wt{G}(k)\ .                                      \e{pw5}
\ee 
Here $J_N(kr)$ is the Bessel function of the first kind and
$k\equiv\sqrt{k_1^2+k_2^2}$. For the scalar field theory the propagator has
the form 
\be
\wt{G}(k)=\frac{1}{k^2+\mu ^2}\ ,                   \e{pw6}
\ee
so that \r{pw5} gives
\bn
G_N(r_1,r_2)&\equiv&\<0|\bar\vf_N(r_1)\vf_N(r_2)|0\>\nbr
&=&\int\,dk\,k\,J_{|N|}(kr_1)J_{|N|}(kr_2)\frac{1}{k^2+\mu ^2}\nbr
&=&\theta(r_1-r_2)K_{\left|N\right|}(\mu r_1)I_{\left|N\right| }(\mu
r_2)
+\theta (r_2-r_1)K_{\left|N\right| }(\mu r_2)I_{\left| N\right| }(\mu
r_1)\ ,                                               \e{pw7}
\en
where $\th(r)$ is the step-function; $I_N,\ K_N$ are the modified Bessel
functions of the first and second kind respectively. 

The principal aim of the ``spherical field theory'' (SQFT) is the development
of a nonperturbative approach to calculations in the standard QFT. We are
interested in UV-behaviour of perturbation expansion in the quantum field
theory on the noncommutative plane (NC-QFT) which we are going to present in
the form similar to the SQFT. Thus to make a comparison, let us first find
out how the UV-divergences of the ordinary (two-dimensional, scalar) field
theory reveal themselves in SQFT. To this aim, we consider the tadpole
diagram depicted in figure~\ref{F:tadp.SQFT}.
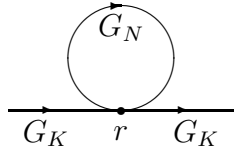
\begin{figure}\centering
\unitlength=1.00mm
\special{em:linewidth 0.4pt}
\linethickness{0.4pt}
\begin{picture}(35.00,24.00)
\put(5.00,10.00){\line(1,0){30.00}}
\put(20.00,10.00){\circle*{1.00}}
\put(20.00,17.00){\circle{14.00}}
\put(10.00,7.00){\makebox(0,0)[cc]{$G_K$}}
\put(30.00,7.00){\makebox(0,0)[cc]{$G_K$}}
\put(20.00,7.00){\makebox(0,0)[cc]{$r$}}
\put(20.00,21.00){\makebox(0,0)[cc]{$G_N$}}
\put(9.00,10.00){\vector(1,0){2.00}}
\put(27.00,10.00){\vector(1,0){2.00}}
\put(19.00,24.00){\vector(1,0){1.00}}
\end{picture}
\caption{\rm The tadpole diagram in the spherical field theory
\e{F:tadp.SQFT}} 
\end{figure}
This diagram is proportional to the factor
\bn
&&\sum_{N=-\infty}^\infty G_N(r,r)=\sum_{N=-\infty}^\infty 
K_{\left| N\right| }(\mu r)I_{\left| N\right| }(\mu r)\nbr
&&\ \ =K_0(\mu r)I_0(\mu r)
+2\sum_{N=1}^\infty 
K_{\left| N\right| }(\mu r)I_{\left| N\right| }(\mu r)\ .  \e{pw8} 
\en
It is seen that the Green function $G_N(r_1,r_2)$ for a fixed partial wave 
is not singular at coinciding arguments: $G_N(r,r)$ has well-defined values
for any $r$ and $N=0,\pm1,\pm2,...\,.$ However, the tadpole diagram is still
divergent: the divergence appears in the summation over the angular momentum
numbers $N$. Indeed, let us for simplicity consider small values of 
$\mu r\ll 1$, so that we can use the asymptotic expressions:
\bnn
\left.I_N(\mu r)\right|_{\mu r\ll 1}&\approx&\frac{1}{N!}
\left(\frac{\mu r}{2}\right)^N\ ,\qq N=0,\pm1,\pm2,...\ ,\nbr
\left.K_N(\mu r)\right|_{\mu r\ll 1}&\approx&\frac{(N-1)!}{2}
\left(\frac{2}{\mu r}\right)^N\ ,\qq |N|\geq1\ ,\nbr
\left.K_0(\mu r)\right|_{\mu r\ll 1}&\approx&\ln\frac{2}{\g\mu r}
\ ,\qq \mbox{($\g$ is the Euler constant)}\ .
\enn
Thus for $\mu r\ll 1$ the tadpole is proportional to
$$
\sum_{N=-\infty}^\infty G_N(r,r)=\ln\frac{2}{\g\mu r}+\sum_{N=1}^\infty
\frac{1}{N}\lra\infty\ ,
$$
so that it is (logarithmically) divergent as it should be. Of course, the
same is true for any values of $\mu r$, though an explicit demonstration of
this fact becomes more involved.

In order to circumvent such a calculation with the special (Bessel) functions,
we can present the action \r{pw3} in a modified form. First, if we are 
interested only in UV-properties of the model, we can drop out the mass term.
However, the massless theory in two dimension has the infra-red divergences
which are also logarithmic for the tadpole diagram 
and may distort the true picture of the UV-behaviour
of the model. Thus, in the massless case we need some IR-regularization 
and to achive it, let us introduce into the (massless) action \r{pw3} the 
additional term of the form
\be
S^{(IR)}=\int_0^\infty\,dr\, \frac{\s^2}{r}\bar\vf_N(r)\vf_N(r)\ .  \e{exp1}
\ee
Now the free action reads as
\be
S^{(0)}=\sum_{N=-\infty}^\infty \int_0^\infty\,dr\left[r
\pad{\bar\vf_N}{r}\pad{\vf_N}{r}+\frac{N^2+\s^2}{r}\bar\vf_N\vf_N
\right]\ ,                                                   \e{exp2}
\ee
and after an introduction of the new coordinate $y$ defined by the relation
\be
y=\ln(\mu r)\ ,\qq -\infty < y <  \infty\ ,        \e{exp3}
\ee
it acquires the simple form of the standard one-dimensional scalar action:
 \be
S_0=\sum_{N=-\infty}^\infty \int_{-\infty}^\infty\,dy\left[
\pad{\bar\vf_N}{y}\pad{\vf_N}{y}+(N^2+\s^2)\bar\vf_N\vf_N\right]\ .\e{exp4}
\ee
The free Green function for this action can be easily found by the use of the 
\fr transform and proves to be the following
\be
G_N(y_1-y_2)=\frac{1}{2M}\er^{-M|y_1-y_2|}\ ,\qq
M\equiv\sqrt{N^2+\s^2}\ ,                            \e{exp4a}
\ee
or, in terms of the initial radial coordinates,
\be
G_N(r_1,r_2)=\frac{1}{2M}\left[\th(r_1-r_2)\left(\frac{r_2}{r_1}\right)^M
+\th(r_2-r_1)\left(\frac{r_1}{r_2}\right)^M\right]\ .      \e{exp5}
\ee
This explicit expression shows immeditely that the tadpole diagram in 
figure~\ref{F:tadp.SQFT} is proportional to the sum
\be
\sum_{N=-\infty}^\infty G_N(r,r)=
\sum_{N=-\infty}^\infty\frac{1}{\sqrt{N^2+\s^2}}\ ,   \e{exp6}
\ee
and, hence, logarithmically divergent. The IR-regularization parameter $\s$ is
inessetial for large $N$ and does not influence on the UV-behaviour of the 
model (as it should be). 

The form \r{exp4} of the action and the UV-behaviour analysis following it
will be useful for us in the noncommutative case as well.
We are going to show that the $\vf^4$-theory on the $q$-plane can be
rewritten as a theory of the partial waves on a one-dimensional
(nonequidistant) lattice with the behaviour in $N$ (angular momentum number)
being even worse, so that UV-divergences of the NC-QFT are even more severe
than those of the usual scalar theory.

\subsection{Quantum field theory on the $q$-plane and its partial wave
decomposition} 
Let us consider the generalization of the two-dimensional scalar field
theory, induced by the noncommutativity \r{2.9} of the space coordinates on
the plane $P^{(2)}_q$. The field action for $\vf(z,\bz)$ can be defined
with the help of the $q$-deformed Haar (invariant) measure \cite{Koelink}.
For a function $F_N(z,\bz)=z^Nf(z\bz)$ on $P^{(2)}_q$ one defines the linear
functional ($q$-integral):
\bn
H_{r_0}[F_N]&\equiv&\int^q\,d^2z\,z^Nf(z\bz)\nbr
&\equiv&\d_{N,0}r^2_0(q^2-1)\sum_{k=-\infty}^\infty 
q^{2k}f(q^{2k}r^2_0)                                    \e{pw9}
\en
($r_0$ labels the nonequivalent representations \r{2.39} of the $q$-deformed
coordinate-momentum algebra). In formula \r{pw9} and in what follows we
assume, for definiteness, that $q^2>1$ (the quite similar construction can be
carried out for $q<1$, \cf \r{corr19}). 
In order that the sum on the right hand side of
\r{pw9} be meaningful, the function $f(z\bz)$ must satisfy an appropriate
conditions at infinity \cite{Koelink}. We shall assume that the set of the
fields on the $q$-plane which we consider below does satisfy this condition. 
Notice that if $N<0$ in \r{pw9}, the integrand can be rewritten as
$\bz^{|N|}f'(z\bz)$ ($f'(z\bz)$ is some modification of the function
$f(z\bz)$).  

Using the $q$-integral \r{pw9}, we can define the action on the $q$-plane as
the straightforward generalization of the usual $\vf^4$-action \r{pw1}:
\bn
S_q&=&\int^q\,d^2z\,\bigg[-\bar\vf(z,\bz)\paq\bpaq\vf(z,\bz)
+\mu^2\bar\vf(z,\bz)\vf(z,\bz)\nbr
&&+\frac{\ld}{2}\bigg(\bar\vf(z,\bz)\vf(z,\bz)\bigg)^2
-j(z,\bz)\bar\vf(z,\bz)-\bar j(z,\bz)\vf(z,\bz)\bigg]\ .      \e{pw10}
\en
Now our aim is to rewrite the action \r{pw10} in the form similar to \r{pw3}
where the integral over radial variables is substituted by the sum \r{pw9}
while the sum over the angular momentum numbers remains in the $q$-deformed
case too. To achieve this, we decompose a field on $P^{(2)}_q$ into terms
with definite eigenvalues of the $q$-deformed angular momentum operator
$J_q$ (\cf \r{2.41}):
\be
\vf(z,\bz)=\sum_{N=-\infty}^\infty z^Nr^{-N}\vf_N(r)
\ ,\qq r^2\equiv z\bz\ .                               \e{pw11}
\ee
Again, as in the case of the expression \r{pw9}, it is worth to notice that
the terms with $N<0$ in the sum \r{pw11} can be rewritten in the form with
positive powers of $\bz$:
\be
z^Nr^{-N}=q^{-|N|(|N|-1)}\bz^{|N|}r^{-|N|}=
q^{-N(N+1)}\bz^{-N}r^{N}\ ,\qq (N<0)\ .                 \e{pw12}
\ee
Here we have used the relation:
\be
z^N\bz^N=q^{N(N-1)}(z\bz)^N\equiv q^{N(N-1)}r^{2N}\ .  \e{pw13}
\ee

The next step is the substitution of the decomposition \r{pw11} into the
action \r{pw10} and then use of the definition \r{pw9} to convert the action
into a lattice one. In order to do this, one needs the
following commutation relations which are derivable from \r{2.9}:
\bn
&&\sqrt{z}\sqrt{\bz}=\sqrt{q}\sqrt{\bz}\sqrt{z}\ ,\qq 
z^Nr^{-N}=q^{-N(N-1/4)}\bz^{-N/2}z^{N/2}\ ,\nbr
&&zr=qrz\ ,\qq \bz r=q^{-1}r\bz\ ,\nbr
&&\paq\sqrt{\bz}=q^{-1}\sqrt{\bz}\paq\ ,\qq
\paq\sqrt{z}=\frac{1}{q^{-1}+1}\frac{1}{\sqrt{z}}+q^{-1}\sqrt{z}\paq\ ,\nbr
&&\bpaq\sqrt{z}=q\sqrt{z}\bpaq\ ,\qq
\bpaq\sqrt{\bz}=\frac{1}{q+1}\frac{1}{\sqrt{\bz}}+q\sqrt{\bz}\bpaq
\ ,                                                   \e{pw14}\\[3mm]
&&\paq z^{N/2}=\frac{[N;q^{-1}]}{q^{-1}+1}z^{N/2-1}+q^{-N}z^{N/2}\paq\ ,\nbr
&&\bpaq\bz^{-N/2}=-\frac{[N;q^{-1}]}{q(q+1)}\bz^{-N/2-1}+q^{-N}
\bz^{-N/2}\bpaq\ ,\nbr
&&\paq\bpaq f(r)=\frac{1}{2+q+q^{-1}}\left[\frac{1}{r}D^{(r)}_{q^{-1}}
rD^{(r)}_{q}\right]f(r)\ .          \nonumber
\en
The Jackson derivatives in the last line are defied as follows:
\bn
D^{(r)}_{q^{-1}}f(r)&=&\frac{f(q^{-1}r)-f(r)}{r(q^{-1}-1)}\ ,\nonumber\\[-1mm]
&&                            \e{corr0}\\[-1mm]
D^{(r)}_{q}f(r)&=&\frac{f(qr)-f(r)}{r(q-1)} \nonumber
\en
(these definitions imply that we are working with the representation where
the operator $r$ is diagonal).

Use of these relations together with \r{pw9} and \r{pw11} allows to present
the action \r{pw10} in the form of the Jackson integral over the radial
variable $r^2$:
\bn
S_q&=&\sum_{N=-\infty}^\infty\int^J\,dr^2\,
\Bigg\{q^{-N(N+1)}\bar\vf_{N}(r^2)\bigg[-\frac{q^3}{(q+1)^2}\D_q\nbr
&&+q^{2(N+1)}\frac{[N;q^{-1}]^2}{(q+1)^2r^2}
-q^{N+1}\frac{[N;q^{-1}]}{q+1)}\Big(q^2 D_{q^2}^{(r^2)}- D_{q^{-2}}^{(r^2)}
\Big)\bigg]\vf_{N}(r^2)\nbr
&&+\mu^2
q^{-N(N+1)}\bar\vf_{N}(r^2)\vf_{N}(r^2)\nbr
&&+\frac{\ld}{2}\sum_{M,K=-\infty}^\infty
q^{2k}q^{-N^2-K^2-M^2-MK+NK+NM-M-K}\nbr
&&\times\bar\vf_{N}(r^2)\vf_{N}(q^{2(N-M)}r^2)
\bar\vf_{N}(q^{2(N-M)}r^2)\vf_{N}(r^2)\nbr
&&+
q^{-N(N+1)}\bar{j}_{N}(r^2)\vf_{N}(r^2)+q^{-N(N+1)}
\bar\vf_{N}(r^2)j_{N}(r^2)\Bigg\}\ .                    \e{pw15m}
\en
Here the Jackson integral for and arbitrary function $f(r^2)$ 
is defined in the standard way (see, \eg \cite{ChaichianDbk}):
\be
\int^J\,dr^2\,f(r^2)\deff r_0^2|q^2-1|\sum_{k=-\infty}^{\infty}q^{2k}
f(q^{2k}r^2_0)\ ,                              \e{corr1}
\ee
and the Jackson derivatives are defined by \r{corr0} and by the following 
similar relations:
\bn
D^{(r^2)}_{q^{-2}}f(r^2)&=&\frac{f(q^{-2}r^{2})-f(r^{2})}{r^{2}(q^{-2}-1)}
\ ,\nonumber\\[-1mm]
&&                            \e{corr2}\\[-1mm]
D^{(r^2)}_{q^{2}}f(r^{2})&=&\frac{f(q^{2}r^{2})-f(r^{2})}{r^{2}(q^{2}-1)}
\ .\nonumber
\en
The radial part $\D_q$ of the $q$-deformed Laplacian reads as 
\be 
\D_q=\frac{1}{r}D^{(r)}_{q^{-1}}rD^{(r^2)}_{q}\ .       \e{corr3}
\ee
Notice that the expression \r{pw15m} for the action $S_q$ in term of the
Jackson integral is equally correct both for the case $q>1$ and for $q<1$. 
For definiteness, we continue to discuss the case $q>1$. The consideration for
the case $q<1$ is essentially the same and we shall present for it only the 
result (\cf \r{corr19}).

Since both the Jackson integral and the Jakson derivatives turn into their
nondeformed (continuous) counterparts in the limit $q\ra1$, it is readily seen
that the action \r{pw15m} becomes in this limit the usual action \r{pw3} 
for the two-dimensional scalar theory in the polar coordinates.

Now we proceed to study the UV-behaviour of the field theory on the
$q$-deformed plane. Therefore, we again, similarly to the nondeformed case 
in the preceding subsection (\cf \r{exp1}), substitue the mass term with the 
IR-regularizing term
\be
S_q^{(IR)}=\sum_{N=-\infty}^\infty\int^J\,dr^2\,
\frac{q^{-N(N+1)}\s^2}{r^2}\bar\vf_{N}(r^2)\vf_{N}(r^2)\ .     \e{corr5}
\ee
The exponential dependence of the fields in \r{pw15m},\r{corr1} 
on the space (discrete) variable $k$ inspires to make the substitution 
similar to that for the nondeformed model (\cf \r{exp3}) and to denote:
\bnn
&&\bar\vf_{Nk}\equiv\bar\vf_N(q^{2k}r_0^2)\ ,\qq
\vf_{Nk}\equiv\vf_N(q^{2k}r_0^2)\ ,\nbr
&&\bar j_{Nk}\equiv\bar j_N(q^{2k}r_0^2)\ ,\qq
j_{Nk}\equiv j_N(q^{2k}r_0^2)\ .                
\enn
In this notation the action \r{pw15m} in which the mass term is subsituted by 
\r{corr5}, acquires the form of an action for infinite 
number of scalar fields on a one-dimensional lattice:
\bn
S_q&=&S_q^{(0)}+S_q^{(int)}+S_q^{(e.s.)}\ ,        \e{corr6}\\[3mm]
S_q^{(0)}&=&\sum_{N=-\infty}^\infty\,A_N\,\sum_{k=-\infty}^\infty
\left[\frac{(\vf_{Nk+1}-\vf_{Nk})^2}{a}
+aM_N^2\bar\vf_{Nk}\vf_{Nk}\right]\ ,    \e{corr7}\\[3mm]
S_q^{(int)}&=&\frac{\ld r_0^2}{2}(q^2-1)
\sum_{M,N,K=-\infty}^\infty\sum_{k=-\infty}^\infty
q^{-N^2-k^2-M^2-MK+NK+NM-M-K}q^{2k}\nbr
&&\phantom{+\frac{\ld r_0^2}{2}(q^2-1)
\sum_{M,N,K=-\infty}^\infty}\times\,\bar\vf_{Mk}\vf_{N(k+M-N)}
\bar\vf_{K(k+M-N)}\vf_{(M-N+K)k}\ ,                   \e{corr8}\\[3mm]
S_q^{(e.s.)}&=&r_0^2(q^2-1)\sum_{N,k=-\infty}^\infty
q^{2k}q^{-N(N+1)}(\bar j_{Nk}\vf_{Nk}+\bar\vf_{Nk}j_{Nk})\ ,\nonumber
\en
where
\bn
&&A_N=q^{-N^2+4}\ ,                     \e{corr9}\\[3mm]
&&M_N^2=\frac{A_N}{R_N}\ ,               \e{corr10}\\[3mm]
&&R_N=q^{-N(N+1)+4}\left(\frac{[N;q]^2}{(q+1)^2}+
\frac{\s^2}{q^4}\right)\ ,              \e{corr11}\\[3mm]
&&a=q^2-1\ .              \e{corr12}
\en
It is obvious that $A_N>0,\ R_N>0$ for all $N=0,\pm1,\pm2,...$. This justifies
the definition \r{corr10} and shows that the quadratic part of the 
Euclidean action 
\r{pw10} is positively defined. The latter fact, in turn, provides that 
the generating functional for Green functions in the model with the action
\r{pw10} (or, in the lattice form, \r{corr6}-\r{corr8}) 
given by the infinite dimensional integral (discrete lattice
analog of the path integral):
\be
\PF[j]=\frac{\int\,\prod_{N,k}\,d\bar\vf_{Nk}\,d\vf_{Nk}\,\exx{-S}}
{\int\,\prod_{N,k}\,d\bar\vf_{Nk}\,d\vf_{Nk}\,
\exx{-\left.S\right|_{j=\bar j=0}}} \ ,                  \e{pw18}
\ee
can be calculated by the perturbation expansion.
A few remarks are in order:
\ben
\item
The fields $\bar\vf_N,\ \vf_N$ at points $q^{2k}r_0,\ k=0,\pm1,\pm2,...$, \ie
the quantities $\bar\vf_{Nk},\ \vf_{Nk}$, can be considered as the creation
and annihilation operators of particles on the quantum plane $P^{(2)}_q$ in
the states \r{2.39}. Thus the field model with the action \r{corr6} 
is the secondary quantized theory of the particles on 
the quantum $q$-plane $P_q^{(2)}$. 
\item
The quadratic part \r{corr7} of the action \r{corr6}  has the standard
form of the lattice scalar theory (\cf \eg \cite{Creutz}), so that we can use
the standard method of the \fr transform in order to
diagonalize it.
\item
As a result of the nonequidistance of the $q$-lattice, the mass term in
\r{pw15m} (the second line) looks as if the model interacts with the external
field, \ie the mass term contains additional factors $q^{2k}$ under the sign
of the sum. This makes diagonalization of the complete quadratic part of the
action \r{pw15m} rather involved problem. 
\item
It is rather striking result that the action on the $q$-plane is not only
lattice-like but also nonlocal, as is seen from the interaction term of the
action in the form \r{corr8}.
\een

The quadratic part $S^{(0)}_q$ of the action can be diagonalized by
performing the Fourier transform
\be
\vf_{Nk} = \int^{\pi /a}_{-\pi /a}\,\frac{dp}{2\pi}\,\er^{\i apk}\, 
\wt{\vf}_N (p)\ .
\ee
Then
\be
S_q^{(0)}=\sum_{N=-\infty}^\infty\,A_N\,\int^{\pi /a}_{-\pi /a}\,
\frac{dp}{2\pi}\,\wt{\bar\vf}_N (p)
\left[\frac{2}{a^2}\Big(1-\cos(ap)\Big) 
+M^2_N \right]\wt{\vf}_N (p)\ ,
\ee
and the free Green function $G_N(k-m)$ has the usual for the lattice 
field theories form (see, \eg \cite{Creutz}):
\be
G_N(k,m)=\int_{-\pi/a}^{\pi/a}\,\frac{dp}{2\pi}\,
\frac{\er^{\i p(k-m)}}{A_N\left[M_N^2+\frac{2}{a^2}\left(1-
\cos(ap)\right)\right]}\ .                         \e{corr14}
\ee

Together with the $\vf^4$-vertex of the action \r{corr8}, 
this free propagator
defines the Feynman rules for the model under consideration which are
depicted in figure~\ref{F:feyn.rules}.
\begin{figure}\centering
\unitlength=1.00mm
\special{em:linewidth 0.4pt}
\linethickness{0.4pt}
\begin{picture}(54.00,33.00)(50.00,00.00)
\put(10.00,10.00){\rule{6.00\unitlength}{2.00\unitlength}}
\put(10.00,11.00){\circle*{2.00}}
\put(16.00,11.00){\circle*{2.00}}
\put(9.00,11.00){\line(-1,-1){4.00}}
\put(17.00,11.00){\line(1,1){4.00}}
\put(17.00,11.00){\line(1,-1){4.00}}
\put(9.00,11.00){\line(-1,1){4.00}}
\put(7.00,9.00){\vector(1,1){1.00}}
\put(19.00,9.00){\vector(1,-1){1.00}}
\put(19.00,13.00){\vector(-1,-1){1.00}}
\put(7.00,13.00){\vector(-1,1){1.00}}
\put(5.00,5.00){\makebox(0,0)[cc]{${\scriptstyle M-N+K}$}}
\put(5.00,17.00){\makebox(0,0)[cc]{${\scriptstyle M}$}}
\put(21.00,5.00){\makebox(0,0)[cc]{${\scriptstyle K}$}}
\put(21.00,17.00){\makebox(0,0)[cc]{${\scriptstyle N}$}}
\put(7.00,11.00){\makebox(0,0)[cc]{${\scriptscriptstyle k}$}}
\put(19.00,11.00){\makebox(0,0)[lc]{${\scriptscriptstyle (k+M-N)}$}}
\put(5.00,30.00){\vector(1,0){8.00}}
\put(13.00,30.00){\line(1,0){8.00}}
\put(13.00,33.00){\makebox(0,0)[cc]{${\scriptstyle N}$}}
\put(64.00,30.00){\makebox(0,0)[lc]{$G_N(k,l)$}}
\put(6.00,28.00){\makebox(0,0)[cc]{${\scriptstyle k}$}}
\put(20.00,28.00){\makebox(0,0)[cc]{${\scriptstyle l}$}}
\put(64.00,11.00){\makebox(0,0)[lc]
{$\frac12\lambda(q^2-1)q^{2k}q^{-N^2-K^2-M^2-MK+NK+MN-M-K}$}}
\put(46.00,30.00){\makebox(0,0)[cc]{$\Leftrightarrow$}}
\put(46.00,11.00){\makebox(0,0)[cc]{$\Leftrightarrow$}}
\end{picture}
\caption{\rm Feynman rules (free propagator and the nonlocal vertex) for the
scalar theory on the noncommutative plane $P^{(2)}_q$\e{F:feyn.rules}} 
\end{figure}  
We shall not carry out detailed perturbative calculations: because of the
nonequidistance of the $q$-lattice such calculations (especially in the case
of nonzero mass) prove to be rather cumbersome. Notice, however, that these
peculiarities of the $q$-lattice seems to be not a difficulty for computer
simulations. In this paper we shall demonstrate only that the UV-divergences
retain in the scalar field theory on the $q$-deformed plane. 
Let us consider the tadpole diagrams presented in figure~\ref{F:tadp}.
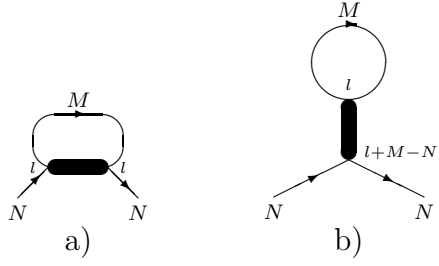
\begin{figure}\centering
\unitlength=1.00mm
\special{em:linewidth 0.4pt}
\linethickness{0.4pt}
\begin{picture}(59.00,34.00)
\put(10.00,12.00){\rule{6.00\unitlength}{2.00\unitlength}}
\put(10.00,13.00){\circle*{2.00}}
\put(16.00,13.00){\circle*{2.00}}
\put(9.00,13.00){\line(-1,-1){4.00}}
\put(17.00,13.00){\line(1,-1){4.00}}
\put(7.00,11.00){\vector(1,1){1.00}}
\put(19.00,11.00){\vector(1,-1){1.00}}
\put(5.00,7.00){\makebox(0,0)[cc]{${\scriptstyle N}$}}
\put(21.00,7.00){\makebox(0,0)[cc]{${\scriptstyle N}$}}
\put(7.00,13.00){\makebox(0,0)[cc]{${\scriptscriptstyle l}$}}
\put(19.00,13.00){\makebox(0,0)[cc]{${\scriptscriptstyle l}$}}
\put(13.00,16.50){\oval(12.00,7.00)[]}
\put(12.00,20.00){\vector(1,0){2.00}}
\put(13.00,22.00){\makebox(0,0)[cc]{${\scriptstyle M}$}}
\put(13.00,3.00){\makebox(0,0)[cc]{a)}}
\put(48.00,15.00){\rule{2.00\unitlength}{6.00\unitlength}}
\put(49.00,15.00){\circle*{2.00}}
\put(49.00,21.00){\circle*{2.00}}
\put(49.00,27.00){\circle{10.00}}
\put(49.00,32.00){\vector(1,0){1.00}}
\put(49.00,34.00){\makebox(0,0)[cc]{${\scriptstyle M}$}}
\put(49.00,24.00){\makebox(0,0)[cc]{${\scriptscriptstyle l}$}}
\put(49.00,14.00){\line(-2,-1){10.00}}
\put(49.00,14.00){\line(2,-1){10.00}}
\put(43.00,11.00){\vector(2,1){2.00}}
\put(53.00,12.00){\vector(2,-1){2.00}}
\put(51.00,15.00){\makebox(0,0)[lc]{${\scriptscriptstyle l+M-N}$}}
\put(39.00,7.00){\makebox(0,0)[cc]{${\scriptstyle N}$}}
\put(59.00,7.00){\makebox(0,0)[cc]{${\scriptstyle N}$}}
\put(49.00,3.00){\makebox(0,0)[cc]{b)}}
\end{picture}
\caption{\rm Two types of the tadpole diagrams in the model under 
consideration\e{F:tadp}} 
\end{figure}  
We confine our consideration to the diagram~\ref{F:tadp}.a because its
analysis is a bit easier than that for the diagram~\ref{F:tadp}.b. Using the
generating functional \r{pw18} and the Feynman rules we find the following 
expression for the tadpole~\ref{F:tadp}.a:
\be
\frac12\ld(q^2-1)q^{-M(M+1)}\sum_{l=-\infty}^\infty 
q^{2l}G_M(k,l)G_M(l,m)\sum_{N=-\infty}^\infty q^{-N(N+1)}G_N(0)\ .\e{pw20}
\ee
The partial wave propagator $G_M(0)$ is finite at the coincident arguments
(similar to the case of the ordinary $\vf^4$-model on a commutative plane,\cf
\r{pw7},\r{pw8} and \r{exp4a},\r{exp5}) and is given by the relation \r{corr14}. 
The simple integration yields
\be
G_M(0)=\frac{1}{\sqrt{R_N(4A_N+a^2R_N)}}\ .              \e{corr15}
\ee 
The divergence again appears in the summation over the
angular momentum numbers: as is seen from \r{pw20}, the tadpole 
contribution is proportional to the factor
\bn
\lefteqn{\sum_{N=-\infty}^\infty q^{-N(N+1)}G_N(0)}\nbr
&&=
\sum_{N=-\infty}^\infty\frac{q^2-1}{q^4\sqrt{[(q^N-1)^2+(q^2-1)^2\s^2/q^4]
[(q^N+1)^2+(q^2-1)^2\s^2/q^4]}}\ .           \e{corr16}
\en
The terms in this series have the following asymptotics:
\bn
\frac{q^2-1}{q^4\sqrt{[(q^N-1)^2+(q^2-1)^2\s^2/q^4]
[(q^N+1)^2+(q^2-1)^2\s^2/q^4]}}&\limar{N}{\infty}&
(q^2-1)q^{-(N+4)}\ ,        \nonumber\\[-1mm]
&&                \e{corr17} \\[-1mm]
\frac{q^2-1}{q^4\sqrt{[(q^N-1)^2+(q^2-1)^2\s^2/q^4]
[(q^N+1)^2+(q^2-1)^2\s^2/q^4]}}&\limar{N}{-\infty}&
\frac{q^2-1}{q^4+(q^2-1)^2\s^2}\ .        \nonumber
\en
The series has the {\it linear} divergence at the lower limit (the
IR-regularization parameter again, as in the nondeformed case, is inessential
in the $N\ra\pm\infty$ limit).
Notice that in the nondeformed limit $q\ra1$ the series \r{corr16} turns into 
the following one
\be
\sum_{N=-\infty}^\infty\frac{1}{\sqrt{N^2+\s^2}}\ ,           \e{corr18}
\ee
and coincides with the nondeformed result (\cf \r{exp6}) (logarithmic
divergence).

Quite similar calculation in the case $q<1$ shows that the tadpole 
diagram~\ref{F:tadp}.a is proportional to the series 
 \be
 \sum_{N=-\infty}^\infty\frac{1-q^2}{q^4\sqrt{[(1-q^N)^2+(1-q^2)^2\s^2/q^4]
[(q^N+1)^2+(1-q^2)^2\s^2/q^4]}}\ ,           \e{corr19}
\ee
which has the linear divergence at the upper limit.

 Thus the perturbation theory for the $\vf^4$-model
on the noncommutative plane $P^{(2)}_q$ with the action constructed with the 
help of the $E_q(2)$--quantum group invariant 
measure contains the UV-divergences
and, hence, can not be considered as a regularization of the usual scalar field
theory on the commutative plane.

\section{Transformation of states on the noncommutative plane $P^{(2)}_q$
induced by the coaction of the quantum Euclidean group $E_q(2)$}
The central problem of this section is to determine
transformation properties of a system on $P^{(2)}_q$ under a coaction of the
quantum group $E_q(2)$. In the subsection 4.1 we shall clarify a general
formulation of this problem and in subsection 4.2 give an explicit answer 
for $E_q(2)$ group. In particular, we shall show that this coaction 
leads to nonlocal transformations of states. 

\subsection{Transformation of states in noncommutative geometry induced by a
quantum group coaction}

Let a quantum group $G_q$ coact on a noncommutative space
$\cX_q$, i.e. there exists the homomorphic map
\be
\delta\,: \ Fun_q(\cX)\lra Fun_q(G)\os Fun_q(\cX)\ ,  \e{1}
\ee
(the algebra $Fun_q(\cX)$ of functions on $\cX_q$ is the configuration
space subalgebra of the algebra of all operators of the given quantum
system). It is natural to say that the system is invariant with respect to 
the quantum group transformations if all the properties of the system are 
independent on the coaction map $\delta$. In other words, the algebra 
$Fun_q(\cX)$ can be realized as the subalgebra of multiple tensor product
$Fun_q(G)\os Fun_q(G)\os...\os Fun_q(\cX)$ and no measurements can
distinguish the description based on the algebras with different numbers
of the factors $Fun_q(G)$.

At first sight, this definition of symmetry transformations may look 
unusual but, in fact, it is a direct generalization of commutative
transformations. Indeed, usual action of a group $G$ of transformations of
a manifold $\cM$ on a function $f\in Fun(\cM)$ is defined by the equality 
\be
T_gf(x):=f(g^{-1}x)\ ,\qquad g\in G,\ x\in\cM\ .       \e{1a}
\ee
The right hand side of this definition can be considered as the function
defined on $G\times\cM$. In other words, the transformations $T$ defines
the map
$$
T\,:\ Fun(\cM)\lra Fun(G)\os Fun(\cM)\ .   $$ 

More customary map $\phi\,:G\os\cM\ra\cM$ is defined for points of the
manifolds, which play the role of the dual set of states for the
commutative algebra of observables (functions) on usual manifolds.
Returning to the transformations with noncommutative parameters, let us 
define the map \cite{Demichev} which is dual to the transformations \r{1} of
observables (operators), \ie
\be
\cS\,: \cH_{G_q}\os\cH_{\cX_q}\lra\cH_{\cX_q}\ ,          \e{2}
\ee
where $\cH_{G_q}$ and $\cH_{\cX_q}$ are the Hilbert spaces of {\it all} 
representations of
the algebras $Fun_q(G)$ and $Fun_q(\cX)$. The
intertwining operator $\cS$ is implicitly defined by the equation
\be
\lgl\delta A|\Psi\os\psi\rgr =\lgl A|\cS (\Psi,\psi )\rgr
\ ,                                                      \e{2a}
\ee
where $\Psi$ is arbitrary vector from $\cH_G$, $\psi$ is arbitrary 
vector from $\cH_\cX$ and $\cS (\Psi,\psi )\in\cH_\cX$ and 
the duality relation $\lgl
A|\psi\rgr\,:\cO\os\cH_\cO\ra\C$ between an operator $A$ from some
algebra $\cO$ and a vector $\psi$ from the Hilbert space $\cH_\cO$ of the
representations of this algebra, is defined by the ordinary mean value of $A$ 
in the state $\psi\,:\ \lgl A|\psi\rgr=\langle\psi|A|\psi\rangle$. 
In fact, the usual
definition \r{1a} of the action of (classical, commutative) transformation
groups in the space of functions on some homogeneous manifold $\cM$ also
has the general form \r{2a}. Indeed, in this case the duality relation
between the algebra $Fun(\cM)$ and states, i.e. points of $\cM$, is
defined as follows
$$
\lgl f|x\rgr=f(x)\ ,\qquad f\in Fun(\cM),\ x\in\cM\ .  $$
The same is true for the group manifold:
$$
\lgl T|g\rgr=T_g\ ,\qquad T\in Fun(G),\ g\in G\ .  $$ 
Thus \r{1a} can be represented in the form
$$
\lgl\delta f|g\os x\rgr = \lgl T\os f|g\os x\rgr
=T_gf(x)                
=\lgl f|\cS (g,x)\rgr=\lgl f|g^{-1}x\rgr=f(g^{-1}x)\ , $$
where the third equality follows from \r{2a} and in this special case 
$\cS (g,x)=g^{-1}x$.

{}From \r{2} it follows that
the matrix elements of the operator $\cS$ in a chosen bases of
$\cH_{G_q}\os\cH_{\cX_q}$ and $\cH_{\cX_q}$ play the role of generalized
Clebsch-Gordan coefficients (GCGC). If the multiple index (set of quantum
numbers) $\{m\}$ defines basis vectors $\psi_{\{m\}}$ of $\cH_{\cX_q}$,
and the set $\{K\}$ defines basis $\Psi_{\{K\}}$ of $\cH_{G_q}$, one can
write 
\be
\psi'=\cS (\Psi_{\{K\}},\psi_{\{m\}})=\sum_{\{l\}} C^{\{K\}}_{\{m\}\{l\}}
\psi_{\{l\}} \ ,                                           \e{3.5}
\ee
where $C^{\{K\}}_{\{m\}\{l\}}$ are the set of GCGC. In this formula the
vector $\psi_{\{m\}}$ is a transformed state on the quantum plane, and the
vector $\Psi_{\{K\}}$ (analog of a point on a grou manifold in the case of
ordinary Lie groups) defines ``parameters'' of the trasformation of
$\psi_{\{m\}}$. 

One can apply analogous consideration to the very quantum group $G_q$
which coacts on itself
\be
M'^i_{\ j}=\Delta M^i_{\ j}=M^i_{\ k}\os M^k_{\ j}\ ,      \e{3.7}
\ee
This leads to the corresponding transformation of vectors in $\cH_{G_q}$
\be
\Psi'=\cS (\Psi_{\{K\}},\Psi_{\{N\}})\equiv
\cS_{\Psi_{\{K\}}}(\Psi_{\{N\}}) = 
\sum_{\{L\}} C^{\{K\}}_{\{N\}\{L\}} \Psi_{\{L\}} \ ,       \e{3.8}
\ee

Two subsequent coactions of the form \r{3.7} induce composition of the
transformations \r{3.8} and general properties of algebra representations
provide its associativity (or, equivalently, this follows from the
coassociativity of Hopf algebras). 
This means that the transformations \r{3.8}
form the {\it semigroup}. The trivial representation 
$\Psi_{\{0\}}\in\cH_{G_q}$ correspond to the identity transformation. 
However, there is no inverse transformation for arbitrary
$\cS_{\Psi_{\{K\}}}$. This means that the transformations \r{3.8},\r{3.5}
do not form a group.

The map $\cS$ satisfies the obvious consistency condition which can be
expressed as a requirement of commutativity of the diagram in 
figure~\ref{cons.rel} (in other words, an equivalence of the different ways
through the diagrams along the arrows).
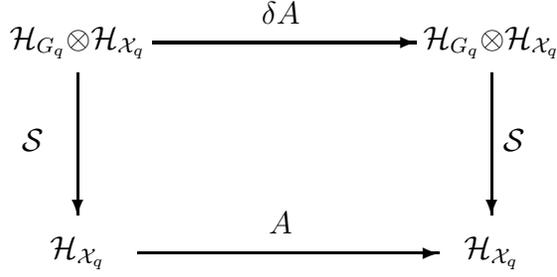
\begin{figure}
\centering
\unitlength=1.00mm
\begin{picture}(68.00,38.00)
\put(10.00,32.00){\makebox(0,0)[cc]{$\cH_{G_q}\os\cH_{\cX_q}$}}
\put(65.00,32.00){\makebox(0,0)[cc]{$\cH_{G_q}\os\cH_{\cX_q}$}}
\put(37.00,36.00){\makebox(0,0)[cc]{$\d A$}}
\put(10.00,28.00){\thicklines\vector(0,-1){19.00}}
\put(10.00,4.00){\makebox(0,0)[cc]{$\cH_{\cX_q}$}}
\put(65.00,28.00){\thicklines\vector(0,-1){19.00}}
\put(65.00,4.00){\makebox(0,0)[cc]{$\cH_{\cX_q}$}}
\put(4.00,19.00){\makebox(0,0)[cc]{$\cS$}}
\put(37.00,8.00){\makebox(0,0)[cc]{$A$}}
\put(68.00,19.00){\makebox(0,0)[cc]{$\cS$}}
\put(20.00,32.00){\thicklines\vector(1,0){35.00}}
\put(18.00,4.00){\thicklines\vector(1,0){40.00}}
\end{picture}\caption{\protect\small The diagrammatic representation 
of the consistency relation for the Generalized Clebsch-Gordan map $\cS$.}
\e{cons.rel}
\end{figure}
For the generalized Clebsch-Gordan coefficients this consistency relation
reads as
\be
\sum_{L,n}C^L_{nr}\< \Psi_L|M^i_{\ j}|\Psi_K\>\< \psi_n|x^j|\psi_m\>
= \sum_{l}C^K_{ml}\< \psi_r|x^i|\psi_l\>\ .                \e{3.10}
\ee
(for the convenience and shortness
we use Dirac bracket notation and drop curly brackets indicating that
$K,m,...$ are multi-indices).  
The equations \r{3.10} must be completed by the normalization conditions
which follow from the normalization of vectors $|K\>$ and $|m\>$.

Notice that the equations \r{3.10} are an analog of the recursion equations
used for determination of ordinary 
Clebsch-Gordan coefficients of $SU(2)$ group. 
However, in the case of quantum groups the problem of explicit solution of
these equation proves to be much more difficult \cite{Demichev}.  

\subsection{Representations of the algebra of functions on $E_q(2)$ and 
transformations of states on $P_q^{(2)}$} 

In this subsection we shall be interested in construction of the
cotransformations of the coordinate subalgebra on a quantum plane and
therefore, for shortness, drop the quantum numbers related to the angular
momentum $J_q$.

According to the discussion in the preceding subsection, a transformation of
a state on a quantum plane depends on a vector from representation space for
the algebra of functions on the group $E_q(2)$. Thus we need explicit
construction of representations of the algebra $E_q(2)$ (\cf \r{2.1}). These
representations have been presented in \cite{VaksmanK}: in slightly different
(more "physical") form they read as 
\bn
&& t=\er^{\se a}\ ,\qq \bt=\er^{\se a^\da}\ ,\qq q^2=\er^{\eta}
\ ,\nonumber\\[-1mm]
&&         \e{4.4}\\[-1mm]
&& v=\er^{\i\f}\er^{\se(a-a^\da)}\ ,\qq \f\in[0,2\pi]\ ,\nonumber
\en
where $a,\ a^\da$ are usual creation and annihilation operators
\be
[a,a^\da]=1\ ,                                     \e{4.5}
\ee
and hence can be represented in any well known way (\eg coordinate, 
Bargmann-Fock, coherent state, infinite matrix representations). The factor
$\er^{\i\f}$ in \r{4.4} is an eigenvalue of the central element $I$ of the 
algebra $E_q(2)$
\be
I=q^{-1}v\bt t^{-1}\ ,                           \e{4.6}
\ee
so that different values of $\f\in[0,2\pi]$ separate different (though
identical) irreducible representations of the $E_q(2)$-algebra. Notice that $I$
is unitary operator $I^\da I=1$, that is why its eigenvalues are parameterized
by $\er^{\i\f}$. 

For an explicit construction we may use the basis of coherent states
\bn
&&a\ket{\z}=\z \ket{\z}\ ,\qq\z\in\C\ ,       \e{4.8}\\[3mm]
&&\<\z'|\z\>=\er^{\bar \z'\z}\ ,\nonumber
\en
so that
\bn
t\ket{\z}&=&\er^{\se\z}\ket{\z}\ ,\nonumber\\[-1mm]
&&\e{4.9}\\[-1mm]
\bt\ket{\z}&=&\ket{\z+\se}\ ,\nbr
v\ket{\z}&=&\er^{-\eta/2+\i\f}\er^{\se \z}\ket{\z-\se}\ ,\qq
\bv=v^{-1}\ .\e{4.10}
\en

The algebra of functions on $P^{(2)}_q$ generated by $\bz,\ z$ has not a
central element and its unique representation has the form \r{4.9} where 
$\bt,\ t$ are substituted by $\bz,\ z$.

According to the general discussion in the preceding subsection, the coaction
of $E_q(2)$ on $P^{(2)}_q$
\bn
\bz\ra\bz'&=&\d\bz\equiv \bv\os\bz+\bt\os\one\ ,\nonumber\\[-1mm]
&&\e{4.11}\\[-1mm]
z\ra z'&=&\d z\equiv v\os z+ t\os\one\ ,\nonumber
\en
induces transformations of states from the representation space
$\cH_{P^{(2)}_q}$ of $P^{(2)}_q$ depending on a state from 
the representation space $\cH_{E_q(2)}$ of the quantum group $E_q(2)$:
\bn
\ket{\xi}\ra\ket{\xi'}&=&\cS\bigg(\ket{\f,\z},\ket{\xi}\bigg)\ ,\e{4.12}\\[3mm]
&&\ket{\f,\z}\in\cH_{E_q(2)}\ ,\qq \ket{\xi},\ket{\xi'}\in\cH_{P^{(2)}_q}\ .
\nonumber
\en
Explicitly this map can be written as follows
\be
\ket{\xi'}=\int\,d\z\,\er^{-\bar\xi\xi}\ket{\xi}\<\xi|\f,\z;\xi'\>\ ,\e{4.14}
\ee
where the generalized Clebsch-Gordan coefficients for the coaction of 
$E_q(2)$ on $P^{(2)}_q$ are denoted as
\be
\<\xi|\f,\z;\xi'\>\equiv\<\xi|\cS|\f,\z;\xi'\>\ .                \e{4.14a} 
\ee

Thus to define transformations properties of the states on $P^{(2)}_q$ we
should calculate the GCGC \r{4.14a}. The consistency relation given by the
commutativity of the diagram in figure~\ref{cons.e2} leads to the set of
equations
\bn
&&\<\xi|\f,\z;\xi'\>=\er^{-\se\z}\<\xi+\se|\f,\z;\xi'\>
-\er^{-\eta/2+\i\f}\er^{\se\xi'}\<\xi|\f,\z-\se;\xi'\>\nbr
&&\<\xi|\f,\z;\xi'\>=\er^{-\se\bxi}\<\xi|\f,\z+\se;\xi'\>
-\er^{-\eta/2-\i\f}\er^{-\se(\z+\bxi}\<\xi|\f,\z+\se;\xi'+\se\>
\ ,\e{4.15}\\[3mm]
&&\<\xi+\se|\f,\z;\xi'\>=\er^{\eta/2}
\er^{\se(\z-\bxi+\xi')}\<\xi|\f,\z-\se;\xi'-\se\>\ ,\nonumber
\en
which must be accompanied by normalization conditions. As we mentioned in the
preceding subsection, it is not easy to solve this equations
straigthforwardly. To circumevent the problem it is helpful to consider the
basis where the primitive elements $\bv,v\in E_q(2)$ are diagonal. 
The point is that a cotransformation for primitive elements has the form of
a cotransformation for an ordinary Lie algebra, 
thus the consistency conditions for them also have a most simple form. 
\begin{figure}
\centering
\unitlength=1.00mm
\begin{picture}(68.00,38.00)
\put(10.00,32.00){\makebox(0,0)[cc]{$\cH_{E_q(2)}\os\cH_{E_q(2)}$}}
\put(65.00,32.00){\makebox(0,0)[cc]{$\cH_{E_q(2)}\os\cH_{E_q(2)}$}}
\put(37.00,36.00){\makebox(0,0)[cc]{$\d v,\d\bt,\d t$}}
\put(10.00,28.00){\thicklines\vector(0,-1){19.00}}
\put(10.00,4.00){\makebox(0,0)[cc]{$\cH_{E_q(2)}$}}
\put(65.00,28.00){\thicklines\vector(0,-1){19.00}}
\put(65.00,4.00){\makebox(0,0)[cc]{$\cH_{E_q(2)}$}}
\put(4.00,19.00){\makebox(0,0)[cc]{$\cS$}}
\put(37.00,8.00){\makebox(0,0)[cc]{$v,\bt,t$}}
\put(68.00,19.00){\makebox(0,0)[cc]{$\cS$}}
\put(23.00,32.00){\thicklines\vector(1,0){29.00}}
\put(18.00,4.00){\thicklines\vector(1,0){40.00}}
\end{picture}\caption{\protect\small The diagrammatic representation 
of the consistency relation for the Generalized Clebsch-Gordan 
map $\cS$ in case
of coaction of the group $E_q(2)$ on itself.}
\e{cons.e2}
\end{figure}

Proceeding in this way, let us construct representation in the basis of the
primitive element $v$. This is easy to do taking into account that in the 
parameterization \r{2.4} the commutation relations for the $E_q(2)$ 
takes the form
\bn
&&[\th,\ro^2]=-\i2\eta\ro^2\ ,                       \e{4.17}\\[3mm]
&& [\th,\nu^2]= 0 ,                                  \e{4.18}
\en
where we have introduced, in analogy with the algebra $P^{(2)}_q$ (\cf 
\r{2.37a}) the operators:
\be
\ro^2\equiv \bt t\ ,\qq \nu^2\equiv \bt t^{-1}\ .       \e{4.18a}
\ee
The only nontrivial commutation relations \r{4.17} is equivalent to 
that for $igl(1,\R)$ Lie algebra (Lie algebra of translations 
and dilatations on a line). Representations
of this algebra are well known (see, \eg \cite{Vilenkin}) and this allows
to write immediately  the required representation with $v=\er^{\i\th}$ being
diagonal:
\bn
vf_\f(x)&=&\er^{\i\f}\er^{-2\i\eta x}f_\f(x)\ ,\nbr
\ro^2f_\f(x)&=&\er^{-\eta}\er^{-\i\pa}f_\f(x)\ ,        \e{4.24}\\[3mm]
\nu^2f_\f(x)&=&\er^{\eta/2}\er^{2\i\eta x}f_\f(x)\ . 
\en
From the form of the operators it is clear that the variable $x$ takes values
on a circle: $x\in [0,2\pi/\eta]$. Thus the functions $f_\f(x)$ are defined
on the circle and form the Hilbert space with the scalar product
\be
\<f_{1\f}(x),f_{2\f}(x)\>=\int_0^{2\pi/\eta}\,dx\,
\bar f_{1\f}(x)f_{2\f}(x)\ .                     \e{4.21}
\ee      

The consistency relations (commutativity of the diagram in
figure~\ref{cons.e2}) for the generators $t,\ \bar t$ result again 
in still rather complicated recursion relations:
\bnn
&&\D t=v\os t+t\os\one\qq\Rightarrow\nbr
&&\left[\er^{\i\f_1}\er^{-2\i\eta(x_1+x_2)}\er^{\i\pa_2/2}+
\er^{-\i\eta x_1}\er^{\i\pa_1/2}-\er^{-\i\eta x}\er^{-\i\pa/2}\right]
\<\f,x|\f_1,x_1;\f_2,x_2\>=0\ ,\nbr
&&\D \bt=\bv\os\bt+\bt\os\one\qq\Rightarrow\nbr
&&\left[\er^{-\i\f_1}\er^{2\i\eta(x_1+x_2)}\er^{\i\pa_2/2}+
\er^{\i\eta x_1}\er^{\i\pa_1/2}-\er^{\i\eta x}\er^{-\i\pa/2}\right]
\<\f,x|\f_1,x_1;\f_2,x_2\>=0\ .
\enn 
But the relation for the primitive element $\bv$ has quite simple form
\bn
&&\D \bv=\bv\os \bv\qq\Rightarrow\nbr
&&\er^{-\i(\f_1-2\eta x_1+\f_2-2\eta x_2)}\<\f,x|\f_1,x_1;\f_2,x_2\>
=\er^{-\i(\f-2\eta x)}\<\f,x|\f_1,x_1;\f_2,x_2\>\ .        \e{4.25}
\en
Here $\<\f,x|\f_1,x_1;\f_2,x_2\>$ denotes GCGC for the algebra $E_q(2)$ in the
realization \r{4.24} and we used the basis $|\f,x\>$ of eigenfunctions of the
operator $\er^{2\i\eta x}$ which are, of course, $\d$-functions on the circle:
\be
|\f,x_0\>=\d^{(S)}(x-x_0)=\frac{1}{2\pi\eta}
\sum_{k=-\infty}^\infty \er^{2\i\eta k(x-x_0)}
\ .                                                    \e{4.26}
\ee
The relation \r{4.25} shows that the eigenvalues $x_1,x_2$ of vectors in the
representations $\f_1$ and $\f_2$ under the tensor product sign and the 
eigenvalue $x$ in an irreducible part $\f$ of the resulting representation are
connected by the relation
\be
-\f_1+2\eta x_1-\f_2+2\eta x_2 =-\f+2\eta x+2\pi n\ ,\qq n=0,1,2,...\e{4.27}
\ee
This is the analog of additivity of the magnetic quantum number in the case of
$su(2)$ Lie algebra: $m=m_1+m_2$ ($m_1,m_2$ are
$J^{(1)}_3,J^{(2)}_3$-eigenvalues of two spins to be summed up and $m$ is an
eigenvalue of $J_3=J^{(1)}_3+J^{(2)}_3$). 

Let us consider the action of the central operator $I$ on the direct product
of two representations
\be
\D I\ket{\f_1,x_1}\ket{\f_2,x_2}\ .                  \e{c1}
\ee
Acting in addition by the intertwining operator $\cS$ and denoting 
$$
\ket{\f_1,x_1;\f_2,x_2}\equiv\ket{\f_1,x_1}\ket{\f_2,x_2}\ ,
$$ 
we have
\bn
\cS\D I\ket{\f_1,x_1;\f_2,x_2}&=&I\cS\ket{\f_1,x_1;\f_2,x_2}\nbr
&=&I\int\,d\f\,dy\,\ket{\f,y}\bra{\f,y}\cS\ket{\f_1,x_1;\f_2,x_2}\nbr
&=&I\int\,d\f\,\ket{\f,y}C(\f;\f_1,x_1;\f_2,x_2)\nbr
&=&\int\,d\f\,\er^{\i\f}\ket{\f,y}C(\f;\f_1,x_1;\f_2,x_2)\ ,  \e{c2}
\en
where according to \r{4.27} we have defined
\be
\bra{\f,y}\cS\ket{\f_1,x_1;\f_2,x_2}\equiv
\d^{(S)}\Big(\f-\f_1-\f_2+2\eta(x_1+x_2-y)\Big)
C(\f;\f_1,x_1;\f_2,x_2)\ .                             \e{c3}
\ee

On the other hand, consider the concrete realization of the operators $I$ and
$\D I$ in $\LLS$ and in the tensor product $\LLS\os\LLS$, respectively. 
To this aim we need an explicit form of $\D\nu^2$ and $\D\ro^2$. Calculation in
the representation \r{4.24} gives the result:
\bn
\D \ro^2&=&\one\os\ro^2+\ro^2\os\one+\bv t\os\bt+\bt v\os t\nbr
&=&\er^{-\eta}\er^{-\i\pa}\bigg[\er^{-\i\wt\pa}
+\er^{\i\wt\pa}+\er^{-\i\f_1}\er^{-\eta/2}\er^{\i\eta x}
+\er^{\i\f_1}\er^{\eta/2}\er^{-\i\eta x}\bigg]\ ,         \e{4.33}\\[3mm]
\D \nu^2&=&\bigg(\bv t^{-1}\os\bt+\nu^2\os\one\bigg)
\bigg(\one\os\one+vt^{-1}\os t\bigg)^{-1}\nbr
&=&\er^{\eta/2}\er^{\i\eta (x+\wt x)}
\frac{1+\er^{-\eta/2}\er^{-\i\f_1}\er^{\i\eta x}\er^{\i\wt\pa}}
{1+\er^{-\eta/2}\er^{\i\f_1}\er^{-\i\eta x}\er^{\i\wt\pa}}\ .   \e{4.34}
\en
Here we used the change of variables
\be
x=x_1+x_2\ ,\qq \wt x=x_1-x_2\ ,                         \e{4.31}
\ee
inspired by the equality \r{4.27}.
Now we can easily calculate the comultiplication for the central operator:
\bn
\D I&=&q^{-1}\D v\D\nu^2\nbr
&=&\er^{\i(\f_1+\f_2)}\er^{-\i\eta x}\er^{\i\eta \wt x}
\frac{1+\er^{-\eta/2}\er^{-\i\f_1}\er^{\i\eta x}\er^{\i\wt\pa}}
{1+\er^{-\eta/2}\er^{\i\f_1}\er^{-\i\eta x}\er^{\i\wt\pa}}\ .   \e{4.35}
\en

In spite of their rather cumbersome form it is easy to check that the
operators $\D \ro^2$ and $\D \nu^2$ indeed satisfy the 
commutation relations of the $E_q(2)$
algebra. In fact, this immediately follows from their general form which can be
written as follows
\bnn
\D \ro^2&=&\const\cdot\er^{-\i\pa}F(x,\wt\pa)\ ,\nbr
\D \nu^2&=&\const\cdot \D I \er^{2\i\eta x}\ 
\enn
(here $F(x,\wt\pa)$ is a function of only $x$ and $\wt\pa$ explicit 
form of which is given in \r{4.33}).

It is clear that if we start from some eigenvector of the operator $\nu^2$
\be
\nu^2\ket{\f_2,x_2}=\er^{\eta/2}\er^{2\i\eta x_2}\ket{\f_2,x_2}\ \e{4.36}
\ee
(\cf \r{4.24}) then, after the comultiplication, we have
\be
\D\nu^2\ket{\f_1,x_1}\ket{\f_2,x_2}=\er^{\eta/2}\er^{-\i(\f_1+\f_2)}
\er^{2\i\eta(x_1+x_2)}\D I\ket{\f_1,x_1}\ket{\f_2,x_2}\ .\e{4.37}
\ee
Since in any irreducible component the central element is proportional to the 
unity operator
\be
\D IP_\f\bigg(\ket{\f_1,x_1}\ket{\f_2,x_2}\bigg)=
\er^{\i\f}P_\f\bigg(\ket{\f_1,x_1}\ket{\f_2,x_2}\bigg)  \e{4.37a}
\ee   
($P_\f$ is a projector onto the irreducible component of the representation
space corresponding to the central element eigenvalue $\er^{\i\f}$), the
relation \r{4.37} shows that for the $\f$-component the eigenvalue of 
$\D\nu^2$ is
\be
\er^{\eta/2}\er^{\i(\f-\f_1-\f_2+2\eta x)}\ ,\qq x=x_1+x_2\ . \e{4.37b}  
\ee
This expression shows how the initial eigenvalue $\er^{\eta/2}\er^{2\i\eta
x_2}$ of the operator $\nu^2$ is transformed under the coaction of $E_q^{(2)}$
with the state $\ket{\f_1,x_1}$ defining the ``parameters'' of the
transformations (\cf \r{3.5}, \r{3.8}). 

The thing which we have to do now is to find the decomposition of an
arbitrary function $f(x_1,x_2)=f(x,\wt x)\in\cH_{E_q(2)}\os\cH_{E_q(2)}$ into
eigenvectors of $\D I$, \ie into irreducible components. To this aim we must
solve the eigenvalue equation
\be
\D Ig_\f(x,\wt x)= \er^{\i\f}g_\f(x,\wt x)\ .                \e{4.40}
\ee
The solution written in terms of the \fr transform
\be
g_\f(x,\wt x)=\sum_{k=-\infty}^\infty \wt g_\f(x,k)\er^{\i k\wt x}\ ,\e{3.40a}
\ee  
has the form
\be
\wt g_\f(x,k)=\er^{\i(\f-\f_1-\f_2+\eta x)k}\er^{\i d(x,k)}\ ,  \e{4.46}
\ee
$d(x,k)$ being defined by the simple recursion relation
\be
d(x,k+1)=d(x,k)-\ld(x,k+1)\ ,\qq d(x,0)=0\ ,           \e{4.45}
\ee
where
\be
\er^{\i\ld(x,k)}\equiv 
\frac{1+\er^{-\eta/2}\er^{-\i\f_1}\er^{\i\eta x}\er^{\eta k}}
{1+\er^{-\eta/2}\er^{\i\f_1}\er^{-\i\eta x}\er^{\eta k}} \ .    \e{4.42}
\ee    
The solution of this recursion is obvious:
\bn
&&d(x,k)=-\sum_{n=1}^k\ld(x,n)\ ,\qq k>0\ ,\nbr
&&d(x,k)=\sum_{n=0}^{k+1}\ld(x,n)\ ,\qq k<0\ ,\nbr
&&\er^{\i d(x,k)}=\prod_{n=1}^k
\frac{1+\er^{-\eta/2}\er^{\i\f_1}\er^{-\i\eta x}\er^{\eta n}}
{1+\er^{-\eta/2}\er^{-\i\f_1}\er^{\i\eta x}\er^{\eta n}} \ ,
\qq k>0\ ,                                              \e{cc1}\\[3mm]
&&\er^{\i d(x,k)}=\prod_{n=1}^k
\frac{1+\er^{-\eta/2}\er^{-\i\f_1}\er^{\i\eta x}\er^{\eta n}}
{1+\er^{-\eta/2}\er^{\i\f_1}\er^{-\i\eta x}\er^{\eta n}} \ ,
\qq k<0\ ,                                               \e{cc2}
\en

The solution \r{4.46}, of course, 
is not square integrable function but a distribution (like
momentum eigenvectors in ordinary Quantum Mechanics). 

Any function $f(x,k)$ can be presented as a \fr integral of the
functions $g_\f(x,k)$
\bn
f(x,k)&=&\frac{1}{2\pi}\int\,d\f\,c(\f)g_\f(x,k)\nbr
&=&\frac{1}{2\pi}\er^{-\i(\f_1+\f_1-\eta x-c(x,k))}
\int\,d\f\,c(\f)g_\f(x,k)\er^{\i\f k}\ ,                 \e{4.49}\\[3mm]
c(\f)&=&\sum_k\er^{\i(\f_1+\f_1-\eta x-c(x,k))}\er^{-\i\f k}f(x,k)\ . \e{4.50}
\en

If we start from eigenvectors of the operator $\nu^2$ (in each component of
the tensor product $\LLS\os\LLS$) 
\bnn
f(x,\wt{x})&\equiv&\ket{\f_1,x_1}\ket{\f_2,x_2}=
\d^{(S)}(x_1-X_1)\d^{(S)}(x_2-X_2)\nbr
&=&\frac{1}{(2\pi\eta)^2}
\sum_{m,n=-\infty}^\infty\er^{\i\eta[m(x-X)+n(\wt{x}-\wt{X})]}\ ,
\qq X=X_1+X_2\,,\ \ \wt{X}=X_1-X_2\ 
\enn
($X_1,X_2$ are the representation variables, while $x_1,x_2$ labels
eigenvalues of the operator $\nu^2$: $\nu^2\d^{(S)}(x_i-X_i)=
\exx{\eta/2+2\i\eta x_i}
\d^{(S)}(x_i-X_i)$), its decomposition over the eigenfunctions $g_\f$ has the
form \r{4.49} with the coefficients:
\be
c(\f,x,\f_1,\f_2)=\d^{(S)}(x-X)\sum_{k=-\infty}^\infty 
\er^{\i(\f_1+\f_2-\f-\eta x)k+\i d(x,k)}\er^{-2\i\eta\wt{X}k}\ . \e{c4}
\ee
Thus 
\be
\D I \ket{\f_1,x_1}\ket{\f_2,x_2}=\frac{1}{2\pi}\int\,d\f\,\er^{\i\f}
c(\f,x,\f_1,\f_2)g_\f(x,\wt{x})\ .                       \e{c5}
\ee
Comparison of \r{c5} and \r{c2} allows to read off the expression for GCGC
of the quantum group $E_q(2)$:
\bn
\bra{\f,y}\cS\ket{\f_1,x_1;\f_2,x_2}&\equiv&\d^{(S)}
\Big(\f-\f_1-\f_2+2\eta(x_1+x_2-y)\Big)C(\f;\f_1,x_1;\f_2,x_2)\ ,\nbr
C(\f;\f_1,x_1;\f_2,x_2)&=&c(\f,x,\f_1,\f_2)\ ,  \e{c6}
\en
and $c(\f,x,\f_1,\f_2)$ is given by \r{c4} and \r{cc1},\r{cc2}.

The corresponding coaction on vectors on $P^{(2)}_q$ have quite the same
form with the only restriction $\f_2=0$. Now we are ready to answer the
question about transformations of the coordinate operators eigenvalues. 
The operator ${u^2}=\bz z^{-1}\in P_q^{(2)}$ (\cf \r{2.37a})
has the eigenvectors similar to those of $\nu^2$:
\be
{u^2}\ket{x_2}=\er^{\eta/2}\er^{2\i\eta x_2}\ket{x_2}\ ,   \e{4.50a}
\ee
After the coaction $\d$ by $E_q(2)$-quantum group, we obtain the 
operator $\d{u^2}$ acting on vectors 
$\ket{\f_1,x_1}\ket{x_2}\in\cH_{E_q(2)}
\os\cH_{P^{(2)}_q}$ and with a structure which is quite similar to $\D\nu^2$ 
(\cf \r{4.37})
\be
\big(\d{u^2}\big)\ket{\f_1,x_1}\ket{x_2}=\er^{\eta/2}\er^{-\i\f_1}
\er^{2\i\eta(x_1+x_2)}\D I\ket{\f_1,x_1}\ket{\f_2,x_2}\ .\e{4.51}
\ee
This implies that the resulting vector can be decomposed into irreducible
parts and, simultaneously, into vectors with definite values of the
coordinate ${u^2}$ as follows:
\be
 \big(\d{u^2}\big)\,\ket{\f_1,x_1}\ket{x_2}=\frac{1}{2\pi}\er^{\eta/2}
 \int\,d\f\,c(\f,x,\f_1,\f_2)
\er^{\i(\f-\f_1+2\eta x)}g_\f(x,\wt x)\ .     \e{4.55}
\ee 

The expression \r{4.55} presents the form of transformation of position
eigenvectors on a quantum plane $P^{(2)}_q$ and shows that the coaction of 
$E_q(2)$ induces nonlocal transformations of the states.

\section{Conclusion}

We have shown that transition to a noncommutative $q$-deformed plane does not
lead to an ultraviolet regularization of the scalar $\vf^4$-quantum field
theory. We start from the firstly quantized theory of quantum particles on
the noncommutative plane. Then we have defined quantum fields depending on
noncommutative coordinates and the field theoretical action using the
quantum analog of the Haar ($E_q(2)$-invariant) measure on the noncommutative
plane. With the help of the partial wave decomposition we have shown that this
quantum field theory can be considered as a second quantization of the
particle theory on the noncommutative plane and that it has
(contrary to the common belief) even more severe ultraviolet divergences than
its counterpart on the usual commutative plane.

We have discussed symmetry transformations on the noncommutative spaces and the
induced transformations of the states. In the case of Lie algebra-like spaces
the coordinates form a tensor operator $\Hx_i\ra
\Hx'_i=M_{ij}\Hx_j+b_i=\hU_g\Hx_i\hU_g^{-1}$ and states of the field system
are transformed by the operator $\hU_g$. We considered the example of such
transformations for the case of noncommutative Euclidean and Minkowski planes
in the preceding paper \cite{CDP98}. In the $q$-deformed case, we have shown
that the quantum group coaction on a coordinate algebra induces nonlocal
transformations of states in the coordinate space. These transformations are
defined by the generalized Clebsch-Gordan coefficients, describing
decomposition of tensor products of representations of algebras of functions
on quantum spaces and representations of the corresponding quantum group. In
other words, the coaction puts in correspondence to a pair of states on a
group algebra $G_q$ and on a quantum space $\cX_q$ some new state on the
quantum space. We have considered such transformations for the case of the
$q$-deformed plane with the $E_q(2)$-symmetry.

\vspace{5mm}

{\bf Acknowledgements} 

The financial support of the Academy of Finland under the Projects No. 44129
is greatly acknowledged. 
A.D.'s work was partially supported also by RFBR-98-02-16769 grant and P.P.'s
work by VEGA project 1/4305/97.

\end{document}